\documentstyle[twoside,epsf]{article}

\catcode`\@=11
\long\def\@makefntext#1{
\protect\noindent \hbox to 3.2pt {\hskip-.9pt  
$^{{\eightrm\@thefnmark}}$\hfil}#1\hfill}		

\def\@makefnmark{\hbox to 0pt{$^{\@thefnmark}$\hss}}	
	
\def\ps@myheadings{\let\@mkboth\@gobbletwo
\def\@oddhead{\hbox{}
\rightmark\hfil\eightrm\thepage}   
\def\@oddfoot{}\def\@evenhead{\eightrm\thepage\hfil
\leftmark\hbox{}}\def\@evenfoot{}
\def\sectionmark##1{}\def\subsectionmark##1{}}

\oddsidemargin=\evensidemargin
\newcounter{sectionc}\newcounter{subsectionc}\newcounter{subsubsectionc}
\renewcommand{\section}[1] {\vspace{12pt}\addtocounter{sectionc}{1} 
\setcounter{subsectionc}{0}\setcounter{subsubsectionc}{0}\noindent 
	{\tenbf\thesectionc. #1}\par\vspace{5pt}}
\renewcommand{\subsection}[1] {\vspace{12pt}\addtocounter{subsectionc}{1} 
	\setcounter{subsubsectionc}{0}\noindent 
	{\bf\thesectionc.\thesubsectionc. {\kern1pt \bfit #1}}\par\vspace{5pt}}
\renewcommand{\subsubsection}[1] {\vspace{12pt}\addtocounter{subsubsectionc}{1}
	\noindent{\tenrm\thesectionc.\thesubsectionc.\thesubsubsectionc.
	{\kern1pt \tenit #1}}\par\vspace{5pt}}
\newcommand{\nonumsection}[1] {\vspace{12pt}\noindent{\tenbf #1}
	\par\vspace{5pt}}

\newcommand{\textlineskip}{\baselineskip=13pt}

\newcommand{\bibit}{\nineit}

\renewenvironment{thebibliography}[1]
	{\frenchspacing
	 \ninerm\baselineskip=11pt
	 \begin{list}{\arabic{enumi}.}
        {\usecounter{enumi}\setlength{\parsep}{0pt}     
	 \setlength{\leftmargin 12.7pt}{\rightmargin 0pt} 
         \setlength{\itemsep}{0pt} \settowidth
	{\labelwidth}{#1.}\sloppy}}{\end{list}}

\newcounter{itemlistc}
\newcounter{romanlistc}
\newcounter{alphlistc}
\newcounter{arabiclistc}

\def\@citex[#1]#2{\if@filesw\immediate\write\@auxout
	{\string\citation{#2}}\fi
\def\@citea{}\@cite{\@for\@citeb:=#2\do
	{\@citea\def\@citea{,}\@ifundefined
	{b@\@citeb}{{\bf ?}\@warning
	{Citation `\@citeb' on page \thepage \space undefined}}
	{\csname b@\@citeb\endcsname}}}{#1}}

\newif\if@cghi
\def\cite{\@cghitrue\@ifnextchar [{\@tempswatrue
	\@citex}{\@tempswafalse\@citex[]}}
\def\citelow{\@cghifalse\@ifnextchar [{\@tempswatrue
	\@citex}{\@tempswafalse\@citex[]}}
\def\@cite#1#2{{$\null^{#1}$\if@tempswa\typeout
	{IJCGA warning: optional citation argument 
	ignored: `#2'} \fi}}

\def\@refcitex[#1]#2{\if@filesw\immediate\write\@auxout
	{\string\citation{#2}}\fi
\def\@citea{}\@refcite{\@for\@citeb:=#2\do
	{\@citea\def\@citea{, }\@ifundefined
	{b@\@citeb}{{\bf ?}\@warning
	{Citation `\@citeb' on page \thepage \space undefined}}
	\hbox{\csname b@\@citeb\endcsname}}}{#1}}

\def\@refcite#1#2{{#1\if@tempswa\typeout
        {IJCGA warning: optional citation argument
	ignored: `#2'} \fi}}

\def\refcite{\@ifnextchar[{\@tempswatrue
	\@refcitex}{\@tempswafalse\@refcitex[]}}

\def\pmb#1{\setbox0=\hbox{#1}
	\kern-.025em\copy0\kern-\wd0
	\kern.05em\copy0\kern-\wd0
	\kern-.025em\raise.0433em\box0}

\def\fnt#1#2{\footnotetext{\kern-.3em
	{$^{\mbox{\scriptsize #1}}$}{#2}}}

\def\runninghead#1#2{\pagestyle{myheadings}
\markboth{{\protect\footnotesize\it{\quad #1}}\hfill}
{\hfill{\protect\footnotesize\it{#2\quad}}}}
\font\tenrm=cmr10
\font\tenit=cmti10 
\font\tenbf=cmbx10
\font\bfit=cmbxti10 at 10pt
\font\ninerm=cmr9
\font\nineit=cmti9

\font\eightrm=cmr8

\def\qed{\hbox{${\vcenter{\vbox{			
   \hrule height 0.4pt\hbox{\vrule width 0.4pt height 6pt
   \kern5pt\vrule width 0.4pt}\hrule height 0.4pt}}}$}}

\begin{document}

\runninghead{Fern\'andez and Rosu, Rev. Mex. F\'{\i}s. {\bf 46 S2} (Nov. 2000) 
153-156}
{Fern\'andez and Rosu $\ldots$}


\normalsize\textlineskip
\thispagestyle{empty}
\setcounter{page}{1}


\vspace*{0.88truein}

\centerline{\bf ON FIRST-ORDER SCALING INTERTWINING IN QUANTUM MECHANICS
     }
\centerline{[\bf  Rev. Mex. F\'{\i}s. 46 S2, (Nov. 2000) 153-156] }
\vspace*{0.025truein}
\vspace*{0.37truein}
\centerline{\footnotesize DAVID J. FERN\'ANDEZ C.}
\vspace*{0.015truein}
\centerline{\footnotesize\it Departamento de F\'{\i}sica, CINVESTAV-IPN,
Apdo Postal 14-740, 07000 M\'exico D.F., Mexico}
\baselineskip=10pt
\vspace*{10pt}
\centerline{\footnotesize HARET C. ROSU}
\vspace*{0.015truein}
\centerline{\footnotesize\it Instituto de F\'{\i}sica,
Universidad de Guanajuato, Apdo Postal E-143, 37150 Le\'on, Gto, Mexico}
\vspace*{0.225truein}

\vspace*{0.21truein}
\noindent
{\bf Abstract.} We generalize the standard first-order intertwining
relationship of
supersymmetric quantum mechanics in order to include simultaneous scaling
transformations in both the original Hamiltonian and the
intertwining operator. It is argued that in this way one can generate
potentials with more interesting spectra than those obtained by means
of the standard first-order intertwining technique and, as an outcome, 
a simple engineering procedure is presented. The harmonic
oscillator potential is used in order to illustrate the previous
statements. Moreover, a matrix representation of the scaled intertwining
relationship is sketched up allowing for higher-dimensional generalizations
in the case of separable potentials.\\ \\
{\bf Resumen.} Generalizamos la relaci\'on de entrelazamiento estandar de
 primer orden de la mec\'anica cu\'antica supersim\'etrica para incluir
 de manera simult\'anea transformaciones de escalamiento tanto en el
 Hamiltoniano original como en el operador de entrelazamiento. Se argumenta
 que en esta forma uno puede generar potenciales con espectros m\'as
 interesantes que aquellos obtenidos por medio de la t\'ecnica de
 entrelazamiento estandar de primer orden y, como un resultado, un
 procedimiento sencillo de ingenieria cu\'antica es presentado. El
 potencial de oscilador arm\'onico es usado para ilustrar las afirmaciones
 anteriores. Mas a\'un, una representaci\'on matricial de la relaci\'on
 de entrelazamiento escalado es bosquejado que permite la generalizaci\'on
 a m\'as de una dimensi\'on en el caso de los potenciales separables.


\textlineskip                  
\vspace*{12pt}                 

\vspace*{1pt}\textlineskip	
\vspace*{-0.5pt}
\noindent


\noindent




\noindent
Factorizations of second order linear one-dimensional (1D) differential
operators are common tools in Witten's supersymmetric quantum
mechanics (SUSYQM) [\refcite{1}], which may be considered as a form of 
Darboux
transformations [\refcite{d}]. They imply either particular solutions 
of Riccati
equations known as superpotentials or the general Riccati solution (RS),
the latter case being first used in physics by Mielnik for the quantum
harmonic oscillator [\refcite{3}]. In recent years, it became clear that
SUSYQM is
not only a form of Darboux transformations but also a particular case in
the more general framework provided by the technique of intertwining
operators [\refcite{intert}], of extensive use in the mathematical 
literature. 
Two (Hamiltonian) operators are said to be intertwined by an operator
$A^{+}$ if the following relationship is fulfilled
\begin{equation}
H_1A^{+}= A^{+}H.
\label{1}
\end{equation}
In SUSYQM $A^{+}$ is a first order differential (factorization) 
operator of the form
\begin{equation}
A^{+}=\frac{1}{\sqrt{2}}\left( -\frac{d}{dx}+\alpha(x)\right),
\end{equation}
leading to the standard Riccati equation for $\alpha(x)$ associated to the
given initial potential
\begin{equation}  \label{R1}
\alpha'(x) + \alpha ^{2}(x)=2(V(x)-{\cal E}),
\end{equation}
where the prime denotes derivative with respect to $x$. The potential
$V_1(x)$ corresponding to $H_1$ is determined according to: 
\begin{equation}  \label{NP1}
V_1(x) = V(x) - \alpha'(x),
\end{equation}
whenever one is able to find a solution of (3) for given $V(x)$ and ${\cal
E}$. The so-called {\it factorization energy} ${\cal E}\in {\bf R}$ plays
a crucial role in generating new solvable potentials from a given one. 
This becomes clear when substituting $\alpha (x) = [\ln\psi(x)]'$ in
(3), which leads to
\begin{equation}  \label{US1}
H\psi(x) = -\frac{1}{2}\psi''(x) + V(x)\psi(x) = {\cal
E}\psi(x) .
\end{equation} 
Although similar to the standard eigenvalue equation for $H$, notice that
$\psi(x)$ in (5) is not necessarily normalizable, but it should not have
zeros in order to avoid supplementary singularities of $V_1(x)$ with
respect to those of $V(x)$. It is well known that for ${\cal E} > E_0$,
where $E_0$ is the ground state energy of $H$, $\psi(x)$ will always have
zeros. However, if ${\cal E} \leq E_0$ it is possible to make $\psi(x)$ to
have no zeros by adjusting the ratio of the two constants in the general
solution of (5), resulting in a physically meaningful $V_1(x)$ as
explained for example by Sukumar [\refcite{3}]. The spectrum of $V_1(x)$
consists of the sequence of levels $E_n$ of the initial Hamiltonian plus a
new `ground state' energy level at ${\cal E}$. The scheme becomes complete
after realizing that $H$ and $H_1$ are factorized in terms of $A$ and
$A^+$ as follows:
\begin{equation}  \label{F1}
H =A A^+ + {\cal E}, \qquad H_1 = A^+A + {\cal E}.
\end{equation}

If now one writes $H_1 = H - f(x)$, the intertwining relationship (1) can
be put in the commutator form as follows
\begin{equation}  \label {com}
[H,A^{+}]=f(x)A^{+},
\end{equation}
and moreover, it is easy to see that Eq.~(4) requires
\begin{equation}
f(x)=\alpha'(x)~,
\end{equation}
that is, in the standard SUSYQM the function $f(x)$ is always the
derivative of the RS $\alpha(x)$, and it will be called the {\em Darboux
potential difference} (DPD) between the Darboux pair of almost isospectral
potentials $V(x)$ and $V_1(x)$, respectively. 

We now look for a more general first-order intertwining relationship and
perform two simultaneous modifications: one related to the Hamiltonian in
which we introduce a ``perturbative" parameter $\epsilon$,
$H_1=(1+\epsilon)H-f(x)$, and the other related to the intertwining
operator (2) which is multiplied (say, to the right) by a unitary operator
$U \equiv e^{i\frac{\lambda}{2}(xp+px)}$ ($U^{+}=U^{-1}$) depending on a
`scaling' real parameter $\lambda$:  
\begin{equation}  \label {2}
A_{\lambda}^{+}=\frac{1}{\sqrt{2}}\left(-D+\alpha(x) \right)U,
\end{equation}
where $D\equiv\frac{d}{dx}$. Multiplying Eq.~(1) by $U^{+}$ to the right
one gets
\begin{equation} \label {mult}
[(1+\epsilon)H-f(x)](-D+\alpha(x))= (-D+\alpha(x))UH U^{+},
\end{equation}
and making equal on both sides of (10) the coefficients of the various
corresponding powers of the derivative operator $D$ leads to
\begin{eqnarray} 
D^3: & e^{-2\lambda}=1+\epsilon , \\ \label {d1}
D^2: &e^{-2\lambda}\alpha(x) = (1+\epsilon)\alpha(x) , \\ \label {d2}
D^1: &(1+\epsilon)\alpha'(x)+(1+\epsilon)V(x)-f(x)=V(e^\lambda x), \\
\label {d3}
D^{0}:& -\frac{(1+\epsilon)}{2}\alpha''(x)+ [(1+\epsilon)V(x) 
-f(x)]\alpha(x)
=\alpha(x)V(e^\lambda x)-V'(e^\lambda x) .  \label{d4}
\end{eqnarray}
One can see that Eqs.~(11) and (12)  are the same, giving the relationship
between the parameters $\epsilon$ and $\lambda$.  After substituting
$e^{-2\lambda}=1+\epsilon$, Eq.~(13)  reads
\begin{equation} \label{11}
f(x)=e^{-2\lambda}\alpha'(x) + e^{-2\lambda}V(x)-V(e^{\lambda}x)~,
\end{equation}
which is a more general form of Eq.~(8). We can see that the original
potential at two different points interfere in the connection between the
DPD $f(x)$ and the RS $\alpha(x)$. This makes a key difference between the
`scaling' intertwining and the standard SUSYQM one, even for the cases when the
potentials are homogeneous functions of degree $d$ for which (15) simplifies 
further to the more local form
$f(x)=s^{-2}\alpha'(x) + (s^{-2}-s^{d})V(x)$,
where $s=e^{\lambda}$. When $d=-2$ one gets 
$f(x)=e^{-2\lambda}\alpha'(x)$ which is the closest one can reach to the
standard SUSYQM. Substituting Eqs.~(11)
and (15) in Eq.~(14) and integrating it we get the Riccati equation
\begin{equation} \label{12}
e^{-2\lambda}\alpha'(x)+e^{-2\lambda}\alpha ^{2}(x) =
2(V(e^\lambda x)-\rm{\cal E})~,
\end{equation}
where, as it will be clear below, it is convenient to choose the
integration constant equal to the factorization energy ${\cal E}$ of (3). 
Now, if one makes the change of variable $y=e^{\lambda}x$ and denotes
$\tilde{\alpha}(y)=e^{-\lambda}\alpha(e^{-\lambda}y)$ one easily gets
\begin{equation} \label{13}
\frac{d\tilde{\alpha}(y)}{dy}+\tilde{\alpha}^{2}(y)=
2(V(y) - {\cal E}).
\end{equation}
One can see that, up to a rescaling of the coordinate, this is the Riccati
equation (3) with a factorization energy ${\cal E}$. Therefore, the same
RS used in the standard intertwining technique (Eqs.~(1) to (8)) can be
used as well in the generalized scaling intertwining technique of
Eqs.(1,9-17)  in order to generate new solvable potentials with known
spectra. The eigenfunctions of $H_2 \equiv e^{2\lambda} H_1= H -
e^{2\lambda} f(x)$ are proportional to the action of $A_\lambda^+$ on the
eigenfunctions of $H$ due to the fact that the two Schr\"odinger
Hamiltonians $H$ and $H_2$ are also intertwined, although now in a
slightly generalized way: 
\begin{equation}
H_2A_\lambda^{+}= e^{2\lambda}A_\lambda^{+}H.
\label{1'}
\end{equation}
The potential of the scaled intertwined Hamiltonian $H_2$ takes the form:
\begin{equation}
V_2(x) = V(x) - e^{2\lambda} f(x) = e^{2\lambda} V(e^\lambda x) -
\alpha'(x) . 
\end{equation}
The corresponding eigenvalues are $\{e^{2\lambda}{\cal E},
e^{2\lambda}E_n\}$, where $e^{2\lambda}{\cal E}$ is the ground state
energy level of $H_2$ associated to the eigenfunction $\psi_{\cal E}(x)
\propto \exp(-\int_0^x\alpha(z) dz)$. The factorizations of $H$ and $H_2$
in terms of $A_\lambda$ and $A_\lambda^+$ become:
\begin{equation}  \label{F2}
H = e^{-2\lambda}A_\lambda A_\lambda^+ + {\cal E}, \qquad H_2
= A_\lambda^+A_\lambda + e^{2\lambda}{\cal E}.
\end{equation}

The generation of new exactly solvable potentials $V_2(x)$ by means of the
scaling intertwining technique depends on the expertness to solve the
Riccati equation (17). In particular, if either we find the general
solution for just an isolated value of the factorization energy ${\cal E}$
or fixing it if we have the solution for ${\cal E}$ in a real interval,
the family of potentials $V_2(x)$ derived by means of this technique would
have the spectrum $\{e^{2\lambda} {\cal E}, e^{2\lambda} E_n\}$, and by
varying $\lambda$ we would be scaling the basic spectrum $\{{\cal E},
E_n\}$ at $\lambda = 0$ generated by means of the standard intertwining
technique when creating a new level at ${\cal E}$ starting from the
initial potential $V(x)$. However, the really interesting procedure is as
follows. Suppose we would solve (17) for ${\cal E}$ belonging to a real
interval $I=[{\cal E}_1, {\cal E}_2]$, where ${\cal E}_1$ and ${\cal E}_2$
have the same sign. By choosing now ${\cal E} = e^{-2\lambda}{\cal E}_0$
with ${\cal E}_0\in I$ fixed and varying $\lambda$ such that always ${\cal
E}\in I$, we would obtain potentials $V_2(x)$ having spectra of the type
$\{{\cal E}_0, (\frac{{{\cal E}}_0}{{\cal E}})  E_n\}$, i.e., changed so
that the excited state levels would be scaled while the ground state
energy level of $H_2$ would be unaffected. This interesting effect is
different of the one achieved by using the standard intertwining
technique, where after solving the Riccati equation (3) for ${\cal E}$
belonging to a real interval, when varying ${\cal E}$ in its corresponding
domain we would change the ground state energy level but maintaining
unaffected the excited state levels. Thus, these two kinds of intertwining
effects, alone or combined, provide us with considerable freedom for
designing solvable potentials with suitable spectra for modeling purposes.

As an example, let us consider the harmonic oscillator potential $V(x) =
x^2/2$. The solution to the Riccati equation (17) for an arbitrary ${\cal
E}<1/2$ is given by (see, e.g., Sukumar and Junker and Roy [3]): 
\begin{equation}
\tilde\alpha (y) = y + \frac{d}{dy}\bigg\{\!\ln \bigg[
{}_1F_1\bigg(\frac{1+2{\cal E}}{4},\frac12;-y^2\bigg) +
2\nu\frac{\Gamma(\frac{3 -
2{\cal E}}{4})}{\Gamma(\frac{1-2{\cal E}}{4})} \, y \,
{}_1F_1\bigg(\frac{3+2{\cal E}}{4},\frac32;-y^2\bigg)
\bigg]\bigg\},
\end{equation}
where, in order to avoid singularities we have that $\vert\nu\vert <1$. 
Thus, the 3-parametric family of potentials (the parameters are $\lambda,
\ {\cal E}, \ \nu$),
\begin{equation}
V_2(x) = e^{4\lambda} \frac{x^2}{2} - e^\lambda \tilde\alpha'(e^\lambda
x),
\end{equation}
with $\tilde\alpha(y)$ given by (21), has spectrum $\{e^{2\lambda}{\cal
E}, e^{2\lambda}(n + 1/2), n=0,1,\dots\}$. By varying $\lambda$ and
maintaining ${\cal E}$ fixed, we will move into members of the family (22) 
attainable after applying the scale transformation onto the potentials
generated by means of the standard intertwining technique, which have
spectrum $\{{\cal E}, E_n = n + 1/2, n=0,1,\dots\}$. On the other hand, if
we take ${\cal E}=e^{-2\lambda}{\cal E}_0\in I_1=(-\infty,0)$, where
${\cal E}_0\in I_1$ is fixed, when varying the scaling parameter $\lambda$
we would be changing ${\cal E}$ so that one will get a two-parameter
family of potentials with spectrum $\{{\cal E}_0, (\frac{{\cal E}_0}{{\cal
E}})(n + 1/2), n=0,1,\dots\}$, i.e., the excited state levels will be
scaled by the factor ${\cal E}_0/{\cal E}$, but the ground state energy
level will remain fixed. A similar treatment can be implemented for ${\cal
E}\in I_2 = (0,1/2)$. A plot of the potentials $V_2(x)$ as a function of
$x$ and ${\cal E}\in I_1$, with ${\cal E}_0 = -1/2$ and $\nu=0$, is shown
in Figure 1. Notice that this selection of ${\cal E}_0$ and $\nu$ ensures
that, up to a displacement of the energy origin, the oscillator potential
is included in this family (for ${\cal E}=-1/2$, i.e., $\lambda = 0$), and
for other values of ${\cal E}$ $V_2(x)$ is symmetric with respect to
$x=0$. For ${\cal E} \in (-1/2,0)$ the potentials $V_2(x)$ have a double
well, and the corresponding spacing for the excited state levels is
expanded by the factor $-1/(2{\cal E})$ ($1$ is the original spacing of
the oscillator potential). The ground state energy level in this case is
fixed at $-1/2$.  On the other hand, for ${\cal E} \in (-\infty,-1/2)$,
the potentials $V_2(x)$ present just a peaked single well centered at
$x=0$, which can be seen as a deformation of the well of the oscillator
potential $x^2/2-1$ needed to maintain fixed the ground state energy level
at $-1/2$. The excited state levels are `squeezed' by the factor
$-1/(2{\cal E})$.

\begin{figure}[htbp]
\begin{minipage}{6truecm}
\hspace*{3truecm}
\epsfxsize=10truecm
\epsfbox{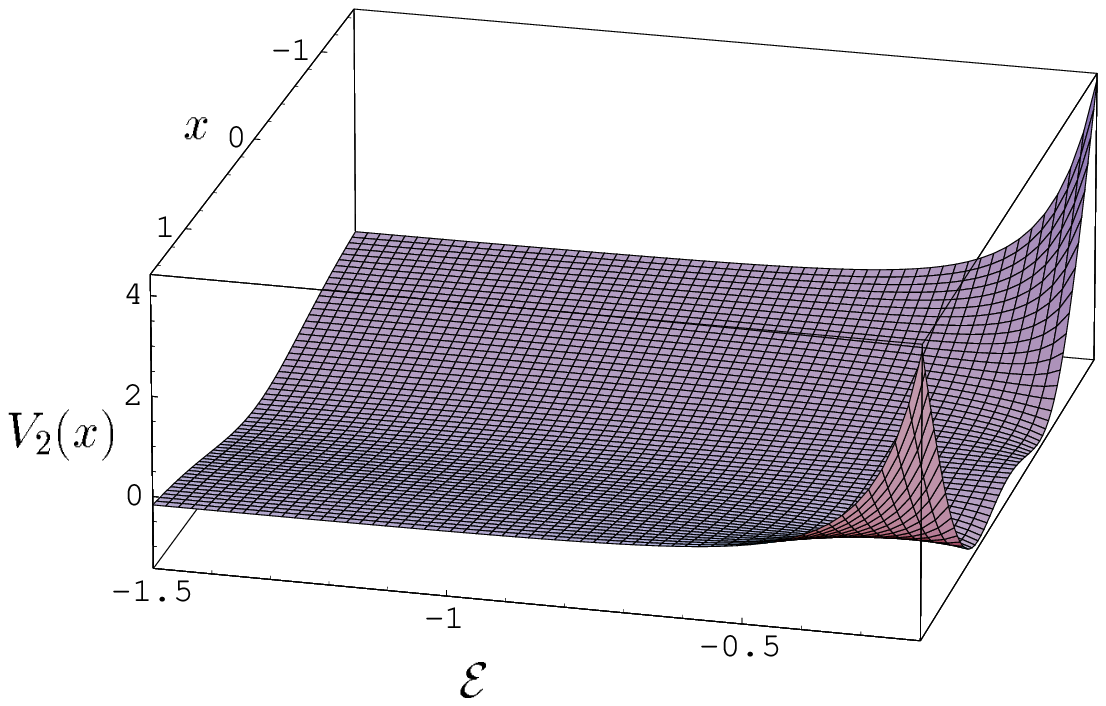}
\end{minipage}
\bigskip
{\caption{\small Plot of the potentials $V_2(x)$ of (22) as a function of
$x$ and ${\cal E}$ when $e^{2\lambda} = -1/(2{\cal E})$ and $\nu=0$. The
oscillator potential $x^2/2-1$ is a member of this family arising for
${\cal E} = -1/2$.}}
\end{figure}

In interesting works, Spiridonov [\refcite{spi}] has used a rather similar
type of scaling intertwining in a different notation (the $q$ one) for
emphasizing the connection between the $q$-deformed calculus
[\refcite{qbooks}] and Shabat's infinite chain of reflectionless
potentials [\refcite{sha}]. In such a treatment the so-called self-similar
potentials have been introduced and characterized, and at the same time
the corresponding spectrum has been determined by means of the algebraic
technique itself. On the other hand, in this paper the orientation has
been different: to take initially a potential with known spectrum in order
to generate a new potential with known and different (in general)
spectrum. Since in the limit $\lambda\rightarrow 0$ we recover the results
derived by means of the standard first order intertwining, we have at hand
an appropriate and interesting generalization of such a technique. For any
other $\lambda$ the new spectrum will not have common elements with the
initial one, and moreover, the above mentioned {\em non-local} influence
of the initial potential on the DPD (see Eq.~(15)) will be at work. This
is one of the interesting {\em novelties} put forth by means of
scaling/deformation within the SUSYQM intertwining procedure.

The generalization of the scaled intertwining formalism to more 
dimensions can be easily performed
for the cases when one deals with separable potentials by means of the
matrix representation [\refcite{rr}]. Thus, for
separable 2D potentials one introduces
${\cal A}^-=\left( \begin{array}{cc}
A & 0 \\
0 & B \end{array} \right ) $  and
${\cal A}^+ =\left (\begin{array}{cc}
A^+ & 0 \\
0 & B^+ \end{array} \right ) $. 
The symbols $A$ and $B$ denote the factorization 
operators for the
first and second coordinate axis, respectively.
The 2$\times$2 matrix intertwining relationship corresponding to (1) reads
\begin{equation} 
{\cal H}_{1}{\cal A}^{+}={\cal A}^{+}{\cal H},
\end{equation} 
where
\begin{equation}
{\cal H}=\pmatrix
{AA^+ + {\cal E}_1 & 0 \cr
0 & BB^+ + {\cal E}_2 \cr}.
\end{equation} 
Corresponding to the function $f(x)$ and the unitary operator $U$
one introduces the following 2$\times$2 diagonal matrices
\begin{equation}
{\cal F}=\pmatrix
{f_1(x) & 0 \cr
0 & f_{2}(y) \cr}= \pmatrix
{\alpha'_{1}(x) & 0 \cr
0 & \alpha'_{2}(y) \cr}
\end{equation}
and
\begin{equation}
{\cal U}=\pmatrix
{e^{i\frac{\lambda _{1}}{2}(xp_{x}+p_{x}x)} & 0 \cr
0 & e^{i\frac{\lambda _{2}}{2}(yp_{y}+p_{y}y)} \cr}~,
\end{equation}
respectively. In (26) the two scaling parameters may or may not be equal. 
By means of these matrices, all the 1D formulas of this 
paper have 2$\times$2 matrix counterparts, though one can immediately see that
in order to maintain the diagonality of matrices one needs separable potentials.
As an example, the matrix form of  
equation (7) reads
$[{\cal H},{\cal A}^{+}]={\cal F}{\cal A}^{+}$.
It is also quite easy to generalize this matrix representation to any number
of even dimensions.

Finally, we draw attention to the fact 
that the scaling intertwining formalism could be
applied to any type of Hamiltonians displaying scaling properties, 
not necessarily quantum ones.
On the other hand, many interesting 2D cases are not separable.
We recall as an example the diamagnetic Kepler problem, i.e., the
Hamiltonian problem of the hydrogen atom in a 
magnetic field [\refcite{hato}], that in atomic units ($\hbar=e=m_{e}=1$)
and for zero angular momentum reads
$H=\frac{1}{2}{\bf p}^2+\frac{1}{8}\gamma ^2(x^2+y^2)-\frac{1}{r}$, 
where $\gamma =B$/(2.35$\times$$10^{5}$T) gives the magnetic field
strength.
In scaled coordinates and momenta, $\tilde{{\bf r}}=\gamma ^{2/3}{\bf r}$ 
and $\tilde{{\bf p}}=\gamma ^{-1/3}{\bf p}$, respectively, this
Hamiltonian can
be written $\tilde{H}=\gamma ^{-2/3}H$ and does not depend on the energy
and 
field strength separately, but only on the scaled 
energy $\tilde{E}=\gamma ^{-2/3}E$. This case, which is also of much
laboratory and astrophysical interest, despite its scaling features,
might be suitable for a more general scaling
intertwining approach with coupled Riccati equations and non-diagonal
matrices. Since here our purposes are merely illustrative
we refer the reader to our recent preprint [\refcite{engi}] 
for more complicated cases.

\nonumsection{Acknowledgements}
\noindent
This work was supported in part by the CONACyT Projects 26329-E and
458100-5-25844E. HCR wishes to thank for the kind
hospitality at the Departamento de F\'{\i}sica, CINVESTAV.



\nonumsection{References}


\end{document}